%% file: wsc24paper.tex
\documentclass{wscpaperproc}
\usepackage{latexsym}
\usepackage{graphicx}
\usepackage{mathptmx}
\usepackage[T1]{fontenc}

\usepackage{amsmath}
\usepackage{amsfonts}
\usepackage{amssymb}
\usepackage{amsbsy}
\usepackage{amsthm}
\usepackage[pdftex,colorlinks=true,urlcolor=blue,citecolor=black,anchorcolor=black,linkcolor=black]{hyperref}

\newtheoremstyle{wsc}% hnamei
{3pt}% hSpace abovei
{3pt}% hSpace belowi
{}% hBody fonti
{}% hIndent amounti1
{\bf}% hTheorem head fontbf
{}% hPunctuation after theorem headi
{.5em}% hSpace after theorem headi2
{}% hTheorem head spec (can be left empty, meaning `normal')i

\theoremstyle{wsc}

\begin{document}

%***************************************************************************
% AUTHOR: AUTHOR NAMES GO HERE
% FORMAT AUTHORS NAMES Like: Author1, Author2 and Author3 (last names)
%
%		You need to change the author listing below!
%               Please list ALL authors using last name only, separate by a comma except
%               for the last author, separate with "and"
%

% setting up general page style
\pagestyle{fancyplain}

% setting up page style of first page
\thispagestyle{plain}
\firstPageHead{}

% setting up running header (authors) of subsequent pages
\chead{\fancyplain{}{\itshape Sandubete-López, Risco-Martín, McMillan, and Besada-Portas}}

% setting up seperation parameters
%\headsep=72pt
\rhead{}
\cfoot{}
\renewcommand{\headrulewidth}{0pt} % (renewcommand needed in fancyhdr to remove top decorative line)
%\headrulewidth=0pt  % ("setlength" needed in fancyheading to remove top decorative line)

\input{wscbib.tex}           % Set up BiBTeX macros

% needed to make the tex document look more like the word counterpart :-(
\setlength{\baselineskip}{12.7pt}

% AUTHOR: Enter the title, all letters in upper case
\title{Methodology for Online Estimation of Rheological Parameters in Polymer Melts Using Deep Learning and Microfluidics}

% AUTHOR: Enter the authors of the article, see end of the example document for further examples
\author{\begin{center}Juan Sandubete-López\textsuperscript{1, 2}, José L. Risco-Martín\textsuperscript{1},  Alexander H. McMillan\textsuperscript{2}, and Eva Besada-Portas\textsuperscript{1}\\
[11pt]
\textsuperscript{1}Dept. of Computer Architecture and Automation, Universidad Complutense de Madrid, Madrid, SPAIN\\
\textsuperscript{2}Microfluidics Innovation Center, Paris, FRANCE\end{center}
}

\maketitle

\vspace{-12pt}

\section*{ABSTRACT}
Microfluidic devices are increasingly used in biological and chemical experiments due to their cost-effectiveness for rheological estimation in fluids. However, these devices often face challenges in terms of accuracy, size, and cost. This study presents a methodology, integrating deep learning, modeling and simulation to enhance the design of microfluidic systems, used to develop an innovative approach for viscosity measurement of polymer melts. We use synthetic data generated from the simulations to train a deep learning model, which then identifies rheological parameters of polymer melts from pressure drop and flow rate measurements in a microfluidic circuit, enabling online estimation of fluid properties. By improving the accuracy and flexibility of microfluidic rheological estimation, our methodology accelerates the design and testing of microfluidic devices, reducing reliance on physical prototypes, and offering significant contributions to the field.

\section{Introduction and related work} \label{sec:intro}

Rheology is the science that studies how materials flow and deform. Among these rheological properties, viscosity is of special interest as it defines how much the fluid will flow when a force is applied. Viscosity estimation is crucial across various applications, from industrial processes like mold injection and 3D printing to medical diagnostics involving bodily fluids \shortcite{shaw:polyrheo,Meek2014,Zhao2023}. The behavior of non-Newtonian fluids, such as polymer melts or gelled propellants, under different pressure gradient presents significant challenges due to the complex dynamics of their viscosities \shortcite{dlrgel2016}.

Traditional methods for measuring viscosity, such as rolling ball, rotational or capillary viscometers, while accurate, are often not suitable for inline measurement and real-time monitoring \cite{shaw:polyrheo}. These standardized methods such as those in \cite{astm2020} typically require discrete liquid samples and are impractical for processes involving continuous or semi-continuous flow of changing fluid compositions. The objective of this study is to develop a methodology that integrates deep learning with microfluidic technology to enable online, real-time estimation of rheological parameters in polymer melts. This approach aims to overcome the limitations of traditional methods by providing a cost-effective, accurate, and flexible solution for continuous monitoring and adjustment of fluid properties.

Recent developments in microfluidic devices are becoming a promising alternative. These devices require minimal fluid volumes and can be directly integrated into larger systems, facilitating real-time rheological estimation \cite{DelGiudice2022}. Various microfluidic designs have already been explored, ranging from devices measuring the deformation of flowing polymers \shortcite{Lee_2005,DelGiudice2020} to systems using embedded sensors to measure pressure changes after a particular flow-path \shortcite{Bie_2019}.

Moreover, in recent years deep learning has been successfully used to model many complex processes in fluid dynamics \shortcite{dlfluids2021} and in microfluidics \shortcite{McIntyre_2022}, as well as in the intersection of rheology and microfluidics to perform parameters identification of non-Newtonian fluids \shortcite{mlvisc2023,Jarujareet_2023}. Nonetheless, estimating the viscosity's parameters from just pressure and flow signals could be convenient for their integration in microfluidic circuits, since these magnitudes are commonly measured. \shortciteN{leastsq2022} do so by combining least squares for the fittings. Their approach requires nevertheless of a particular stepped input signal, not being adequate for online estimation. The use of deep learning could help reducing the dependency of this restricted input sequences, allowing for estimating rheological properties, online, from pressure and flow rate signals.

One aspect that makes difficult the development of these systems is the fact that deep learning models requires of large training datasets. Simulating Newtonian and non-Newtonian fluids is often a computationally expensive and time-consuming task, which requires both expertise and resources because it is commonly done by means of Computational Fluid Dynamics (CFD, \shortciteNP{GalindoRosales2010}). For microfluidics, the simulations might be performed applying a finite-volume method \shortcite{Keslerov_2010,ferziger2019cfd} or the Lattice-Boltzmann method \shortcite{kruger2017lattice}. Higher levels of abstraction are possible, using the geometry of the systems and their symmetries to approximate the microfluidic system as two-dimensional \shortcite{Boulais_2023} or one-dimensional \shortcite{elecanalog2012} problems. This allows to trade off the accuracy and complexity of the simulations. In this regard, \shortciteN{Takken_2024} have recently proposed a hybrid simulation framework in which high levels of abstraction are used to speed-up the simulations through analytic expressions (where the flow path is abstracted to 1D), using CFD only in cases where this expressions are not available (e.g. junctions of fluidic paths). 

Our methodology employs these principles of microfluidics to reduce the complexity of simulating generalized non-Newtonian fluids behaviors approximating the system as one-dimensional. By modeling the microfluidic system as a network of hydraulic resistances and capacitances, we can emulate the flow characteristics of various fluids under different conditions with reduced computational overhead. This model generates synthetic data used to train a deep neural network, an architecture containing Recurrent Neural Networks (RNN), which allows to discover time dependencies on the input data. This deep learning model is trained to predict the fluid's rheological parameters based on observed pressure changes and flow rates.

The main contributions of this work are: a methodology which can be applied to model, simulate and develop microfluidic systems flowing non-Newtonian fluids and integrated with deep learning, and an innovative approach to viscosity estimation for generalized Newtonian fluids, and more concretely, polymer melts, from commonly measured magnitudes such as pressure and flow signals. By providing a robust simulation framework, this work contributes to the broader field of fluid simulation and modeling, offering a practical tool for researchers and engineers working with complex fluid systems.

Examples of potential applications might range from the development of deep neural networks for the estimation of viscosity's parameters (shown in this paper), to chemical process AI-based control in microfluidics, passing through cell-culture anomaly detection within complex fluids.

This paper is organized as follows. Section \ref{s:bckg} describes the theoretical microfluidic background underpinning our methodology. Section \ref{s:method} details our methodology, including the exposition of our microfluidic circuit design, of the physical model used for simulation, and of the architecture of the neural network model used to estimate parameters of the viscosity model. Section \ref{s:ear} presents the conducted experiments and the obtained results. Finally, Section \ref{s:cafw} draws the conclusions and presents some future lines of research.

\section{BACKGROUND} \label{s:bckg}

This section introduces the required background on two areas supporting our work: generalized Newtonian fluid models, and microfluidics systems modeling.

The viscosity of a fluid relates its rate of displacement or shear rate, usually denoted $\dot{\gamma}$, for an imposed shear stress. This shear stress depends of how the material reacts to an applied force under some conditions. For Newtonian fluids, the ratio between these two magnitudes is independent of the applied shear stress, while in non-Newtonian fluids this does not always hold. Many different models have been proposed, although the most commonly used is the power-law fluid model \shortcite{shaw:polyrheo}, 
\begin{equation} \label{eq:pwlf}
\eta = \eta_{0} (\dot{\gamma})^{n-1}, 
\end{equation}
where, $\eta_{0}$ stands for the zero shear rate viscosity, $\dot{\gamma}$ for the aforementioned shear rate, and $n$ for its power index. For $n=1$, shear rate and viscosity are perfectly independent, which is the behavior of an ideal Newtonian fluid. This model is used by \shortciteN{Srivastava2006} to semi-analytically approximate the solution to the flow of a non-Newtonian fluid through a hydraulic resistance of rectangular cross-section.

Analytic and semi-analytic expressions have been derived for the streamlined capillary geometries commonly used in microfluidics. These expressions are obtained by considering one-dimensional or two-dimensional approximations to the problem \shortcite{mortensen2004}. They are used to speed up the simulations of these systems, accelerating the design process, for example, by simulating the system as analogous electrical problems, i.e., 1D approximation of the microfluidic circuits \shortcite{elecanalog2012}.

In these analogous electrical problems, an equivalent Ohm's law \shortcite{tipler2004} is defined for Newtonian fluids under laminar flow. The equations of Hagen-Poiseuille \cite{Anna_2013} allows for solving microfluidic circuits in an analogous manner to electrical circuits. In particular, the flow rate, $Q_{v}$, through a hydraulic resistance can be estimated as
\begin{equation}\label{eq:hydOhm}Q_{v} = \frac{\Delta P }{R_{hyd}},\end{equation}
where $R_{hyd}$ is the \textit{hydraulic resistance} value, and $\Delta P$ is the pressure drop over it. Here, the subscript $v$ of $Q_{v}$ is used to denote that the flow is governed by viscous forces. If the circuit has only one branch with multiple hydraulic resistances interconnected in series, then the total resistance of the circuit can be estimated as $R_{hyd}=R_1 + R_2 + ...$. The total flow rate passing through the branch can be calculated using equation \eqref{eq:hydOhm} over the total resistance.

Lastly, and continuing with the analogy, a hydraulic $RC$ circuit can be defined by adding a hydraulic capacitance, $C$, to the circuit. Its transient response will be then governed by a time constant $\tau = R_{hyd}C$. The relation between volume, $V$, and capacitance $C$ is simply $C = K_{fluid} V$, where $K_{fluid}$ depends on the fluid, being $K_{fluid}=1.0$ for air.

\section{METHODOLOGY} \label{s:method}

\begin{figure}[b]
    \centering
    \includegraphics[width=1.0\textwidth]{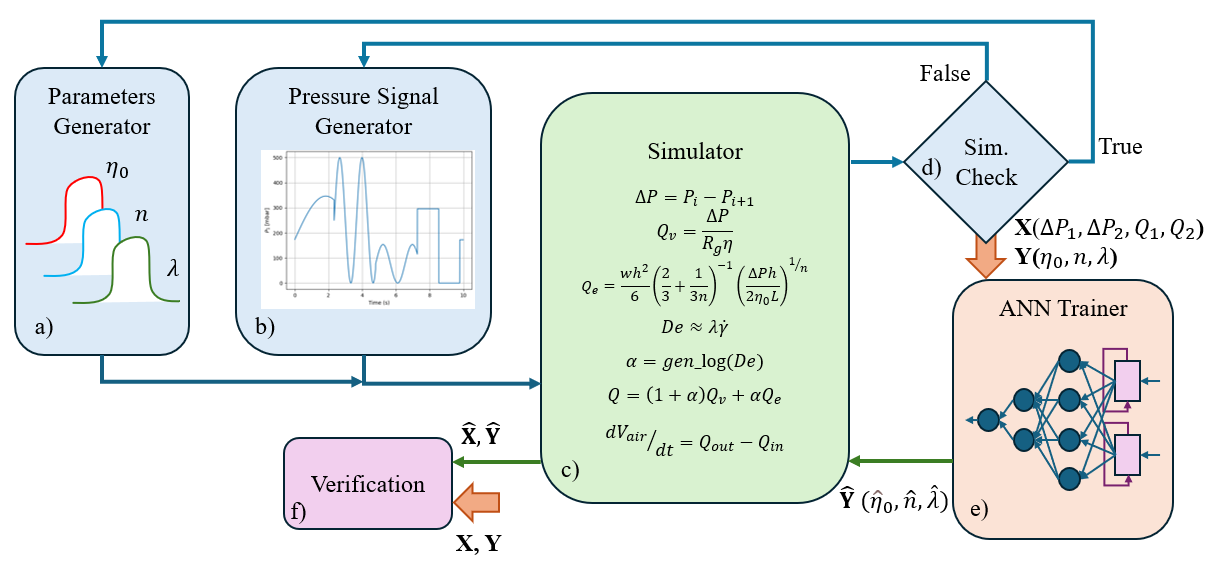}
    \caption{Methodology diagram. The elements in blue, a), b) and d), operate together with the simulator to generate a dataset, \textit{X, Y}, with which the deep learning model is trained at block e). The simulator uses the predictions of the deep learning model to run simulations, producing an estimation dataset, $\hat{X}, \hat{Y}$, which is compared at f) with the original dataset to verify the methodology.}
    \label{fig:MethDiag}
\end{figure}

Figure \ref{fig:MethDiag} shows the overall work-flow of the designed and implemented methodology, which is explained in this section. The aim of the methodology is to generate data for training a RNN model to estimate the rheological properties of a fluid.

First, a set of values of the viscosity model's parameters is sampled from random uniform distributions within defined ranges. For this set, a random sequence of pressure signals is generated to be used as input to the microfluidic circuit simulator together with the sampled parameters, outputting signals of pressure drop along the resistances, as well as input and output flows to the capacitance. Simulations that do not contain significant information for the corresponding parameters are rejected, requiring the generation of a new random sequence of signals for the same set of parameters. This way, for each simulation, four output signals and one set of parameters of the viscosity model are collected.

Second, the generated dataset is used to perform a supervised training of the RNN model with the purpose of identifying the viscosity's parameters corresponding to each fluid from the signals produced by the simulator. This way, the model is trained for a number of epochs using mini-batches where each mini-batch is composed of sixteen simulations. Each simulation is made up of the four signals (two pressure drop signals, and two flow-rate signals) associated to the set of viscosity model's parameters that have been used to produce them. The RNN model receives these four signals, sample by sample, four values in parallel (one by input signal), outputting one prediction once the whole simulation has passed through.

Last, the simulator is again used to verify the operation of the trained artificial neural network. The input signals from the test dataset are used to test the prediction quality of the RNN model. For each simulation, the RNN will predict the corresponding set of parameters. These estimated parameters are then simulated and the output four signals of pressure drop and flow rate are compared to the ones produced with the correct set of parameters of viscosity. The difference is evaluated to verify that, for each simulation, if the predicted parameters are different, the output of the simulation is different as well. This indicator is used as guide for the discovery of the right setup to estimate the parameters of the viscosity model.

In the following subsections, the proposed microfluidic circuit design is described along with the corresponding physical model. The simulation process is then introduced, followed by the description of the proposed artificial neural network architecture. Lastly, the used verification process is detailed.

\subsection{Design and Modeling of the Microfluidic Circuit}
\begin{figure}[t]
    \centering
    \includegraphics[width=0.7\textwidth]{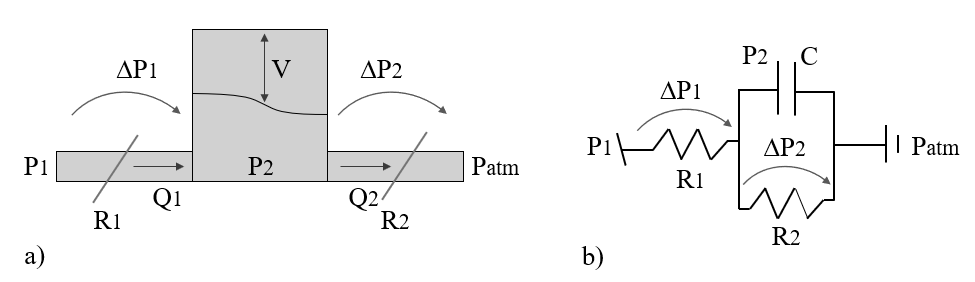}
    \caption{Simulated microfluidic circuit. a) shows the fluidic system diagram composed by one input hydraulic resistance, $R_1$, one capacitance characterized by its volume, $V$, and one output hydraulic resistance, $R_2$. b) shows the analogous electrical circuit.}
    \label{fig:SysDiag}
\end{figure}

Microfluidic systems development is pursued for many reasons (e.g. increased surface-volume ratio or  low-waste applications, \citeNP{tabeling2005introduction}), however, the most interesting feature of microfluidics related to this work is the reduction of complexity of the system under laminar flow. According to Hagen-Poiseuille's equations \cite{Anna_2013}, the laminar flow  of a Newtonian liquid passing through a capillary can also be analytically characterized for simple cross-sectional geometries. The expression depends of the geometry and dimensions of the capillary, the dynamic viscosity of the fluid, $\eta$, and the applied pressure drop along the capillary \cite{tabeling2005introduction}. In particular, for a microfluidic resistance of rectangular cross-section of width, $w$, height, $h$, and length, $L$ (when $h \ll w$, and $h, w \ll L$) becomes
\begin{equation} \label{eq:HP}
Q_{v, i} = \frac{ wh^3 \Delta P_i}{12L \eta} \left(1 - 0.63 \frac{h}{w}\right)= 
\frac{\Delta P_i}{R_{g,i} \eta},
\end{equation}
where $R_{g,i}=12L/(wh^3(1 - 0.63 \frac{h}{w}))$ stands for the contribution of all the purely geometrical terms to the $i-th$ hydraulic resistance (i.e. $R_{i}=R_{g,i} \eta$). The subscripts, $i$, in terms $h, w,$ and $L$ have been omitted for simplicity in equation \eqref{eq:HP}.

Figure \ref{fig:SysDiag} shows the equivalence between the designed microfluidic circuit and the corresponding ideal electrical circuit, which can be modeled by one resistance in series connected to another resistance with a capacitor in parallel, with $R_1 = R_2$. For Newtonian fluids, the rate at which the capacitance changes its internal pressure is given by a time constant, $\tau$, so resolving by applying the Thevenin's theorem: $\tau = R_{th} C = (R_1+R_2) C$. In this model, the hydraulic resistance corresponding to the tubing which connects the elements (i.e., from the reservoir to the $R_1$, and from $R_2$ to the waste) and the connectors (hydraulic interfaces between elements) have been neglected since the characteristic dimensions of the cross-section of the capillary are usually around two orders of magnitude above, and the hydraulic resistance decreases with the power of three over these dimension. 

In this hydraulic $RC$ circuit in which a Newtonian fluid flows, if the values of $R_{g}$ and $C$, are known, the viscosity can straightaway calculated as $\eta = \tau / (R_{g}C)$ after measuring the circuit time constant $\tau$. The introduced capacitance is thus important because it allows to extract information of the fluid from the transient response.

As it can be observed in Figure \ref{fig:SysDiag}, an extra resistance (called $R_1$ hereafter) is included before the capacitance in the hydraulic circuit, or before the $RC$ in the electrical analogy. This input resistance has the purpose of limiting the maximum flow rate to the capacitance.  Following with the electrical analogy, it will short-circuit for fast changes in the input signal, leading to high flow-rates that could move the regime out of laminar flow. The value of this new resistance, although arbitrary, is set to be equal to the resistance of the $RC$ part (called $R_2$ hereafter) for simplicity. Estimates of their values can be calculated from commercially available microfluidic resistance chips with $h=20 \mu m$, $w=800\mu m$, and $L=25mm$, which leads to $R_{g1} = R_{g2} = 9.53e16$ $m^{-3}$.

In polymer melts and other polymeric dissolutions, equation \eqref{eq:HP} holds while the polymer chains recovery time is smaller than the characteristic time of the flow, because the elastic forces do not have a significant influence over the dynamics of the fluid. The dimensionless Deborah number can be used to estimate when these elastic forces must be taken into account \cite{Anna_2013} and are computed according to 
\begin{equation} \label{eq:De_eq}
De_i \approx \lambda \dot{\gamma}_i = \frac{6 \lambda Q_{v,i}}{wh^2},
\end{equation}
where $\lambda$ is the relaxation time of the fluid, and $\dot{\gamma}_i$ the shear rate. Being the rightmost part of the equation the corresponding solution at the wall of the $i-th$ capillary of rectangular cross-section. 

For $De > 1/2$, the chains are elongated with the flow and the elasticity of the fluid plays a significant role on the overall observed behavior which cannot be neglected. In this case,  semi-analytic equation \eqref{eq:NNHP} obtained by \shortciteN{Srivastava2006} can be used to approximate the flow rate through the given $i-th$ rectangular hydraulic resistance,
\begin{equation} \label{eq:NNHP}
Q_{e,i} \approx \frac{wh^2}{6} \left(\frac{2}{3} + \frac{1}{3n} \right)^{-1}
\left(\frac{h\Delta P_i }{2 \eta_0 L}\right)^{1/n},
\end{equation}
where $\eta_0$ stands for the zero-shear-rate viscosity (which corresponds to $\eta$ in Newtonian regime) and $n$ for the power-law index. The subscript $e$ in $Q_{e,i}$ is used to denote the flow under influence of elastic forces.

In order to produce a set of equations that allow smooth transitions between the Newtonian and non-Newtonian behavior, the set of expressions is complemented with a generalized logistic function that receives as input the value of $De$ and produces as output the value of $\alpha$, which is used to weight equation \eqref{eq:HP} and equation \eqref{eq:NNHP} during the transition according to:

\begin{equation} \label{eq:Qf}
Q_i = (1-\alpha)Q_{v,i} + \alpha Q_{e,i}
\end{equation}

The flow rate, $Q_i$, through each resistance, $R_1$ and $R_2$, might be different during the transient response, while the capacitance, $C$, charges or discharges fluid. Lastly, the only state variable considered in our microfluidic simulator is the charging of the capacitance, $C$. Modeled as a capacitance of the compressed air, the evolution of the volume can be calculated as the balance of fluid flow experienced by the capacitor:
\begin{equation} \label{eq:Vdot}
\frac{dV}{dt} = Q_{2} - Q_{1},
\end{equation}
where $Q_{1}$ is the flow going into the capacitance, $C$, and $Q_{2}$ is the flow going out of it. Each of these flows will be calculated according to equation \eqref{eq:Qf}.

The rate of change of volume from equation \eqref{eq:Vdot}, integrated, can be related to the pressure by means of Boyle-Mariotte formula \cite{tipler2004}. For the first hydraulic resistance, the expression will become:
\begin{equation} \label{eq:incP}
\Delta P_1 = P_{1} - P_2 = P_{1} - P_o \frac{V_{max}}{V},
\end{equation}
where $V_{max}$ is the maximum volume of air stored in the capacitance, $P_o$ is its initial internal pressure, and $V$ is the volume of air stored inside, calculated by integrating $dV/dt$, and $P_1$, $P_2$ are the pressures at the input and the output of the first hydraulic resistance, $R_1$. Equivalently, the pressure drop along $R_2$ is calculated as $\Delta P_2 = P_{2} - P_3$.

\subsection{Simulation}
\begin{figure}[t]
    \centering
    \includegraphics[width=0.85\textwidth]{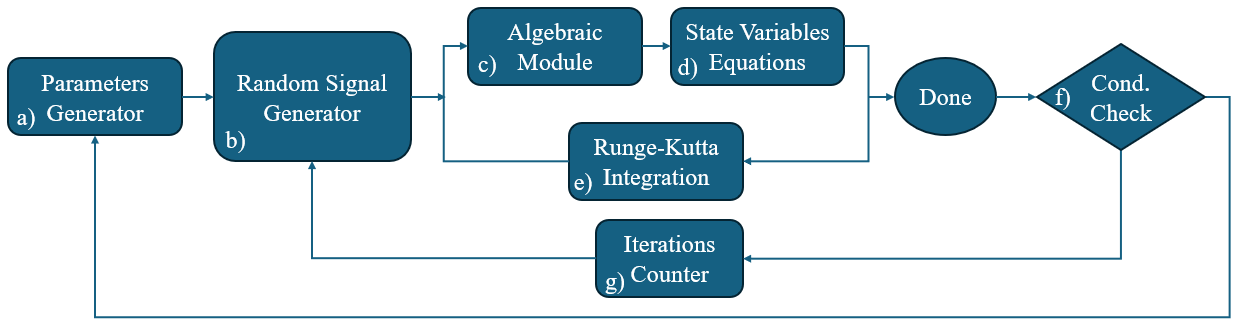}
    \caption{Diagram showing the simulation flow. a) generates set of parameters for which block b) creates a random input pressure signal. In the simulation loop, c) calculates all the algebraic expressions and d) resolves the variation of volume in the capacitance, integrated by e). If the conditions of f) don't match, a new signal is generated and the loop is run again. Otherwise, a new set of parameters is created and the process starts again. }
    \label{fig:sim_diag}
\end{figure}
To generate the training and testing datasets for the RNN, simulations are run following the scheme presented in Figure \ref{fig:sim_diag}. The first block generates a set of random values for the parameters $\eta_0$, $n$ and $\lambda$, from their corresponding uniform distributions within the ranges defined in Table \ref{tab:ranges}. These ranges has been selected according to the selected commercially-available hydraulic resistances to be able to observe Newtonian and non-Newtonian behaviors within experiments of 12.5 seconds. The timescale of the tool can be tuned by changing the time duration of the experiments, the values of the hydraulic resistances or the capacitance of the microfluidic circuit in order to  study fluids of different degree of viscosity with this setup. 

\begin{table}[b]
\centering
\caption{Ranges of the uniform distribution for each studied parameter of the fluid.\label{tab:ranges}}
\begin{tabular}{rll}
\hline
Parameter & Minimum & Maximum\\ \hline
$\eta_0$ & $1.0e-3$ & $1.6e-3$\\
$n$ & 0.90 & 1.1 \\
$\lambda$ & $1.0e-4$ & $5.0e-4$\\
\hline
\end{tabular}
\end{table}

As Table \ref{tab:ranges} shows, shear-thinning (which happens when $n<1.0$) and shear-thickening (when $n>1.0$) are allowed behaviors for the fluid. Nevertheless the range of $n$  remains close to 1.0, because the tubing and connections have been neglected in the current equivalent circuit. This is, high shear-thinning fluids, like xanthan solutions \shortcite{Mrokowska_2019} or gelled propellants with $n < 0.5$ \shortcite{Caldas_Pinto_2019}, will be less accurately simulated under non-Newtonian regime the smaller their behavior exponent is, because higher viscosities will be observed at the tubing, increasing their resistance to flow. And, oppositely, shear-thickening fluids simulations will be more accurate the higher $n$ is.

After the random values of the parameters are generated, a second block generates random sequences of sines, steps and ramps. The reason behind the randomness of the input signals is to increase the variance in the dataset to avoid having to  impose patterns in the input signals for estimating the parameters, minimizing the interference with the user's work in a real setup.

Next, in order to generate datasets with valid information, a type of rejection sampling scheme is applied. The process starts using the sampled values of the parameters and of the reference inputs to run the simulations, and continues checking if the simulations output signal fulfill the conditions required by the \textit{conditions check} block. 

The simulation is performed such that, at each instant, $k$, $P_1$ is imposed from the input pressure signal, then equation \eqref{eq:incP} is used to calculate the pressure drop over $R_1$. This is used in equation \eqref{eq:HP} to calculate the Newtonian flow-rate, $Q_{v,1}$, and in equation \eqref{eq:NNHP}, to calculate the non-Newtonian flow rate, $Q_{e,1}$, through $R_1$. $Q_{v,1}$, is then used as input to calculate $De_1$. From this number, the parameter $\alpha_1$ is obtained and used to weight $Q_{v,1}$ and $Q_{e,1}$ for the final $Q_1$ calculation. The process is repeated for $R_2$, obtaining the corresponding $Q_2$. Finally, the variation rate of the volume at the capacitance is calculated as the difference between $Q_2$ and $Q_1$.

The condition block checks if  the output signal contains information of Newtonian and non-Newtonian behavior of the fluid. In particular, if  $\Bar{\alpha} = \{\alpha_0, \alpha_1, ..., \alpha_N \}$ is the sequence of values of $\alpha$ that establishes the regime of the flow at equation \eqref{eq:Qf} at different instant $k \in \{0, N\}$, the condition block checks if
\begin{equation} \label{eq:cond1}
|\Bar{\alpha_v}| = |\{ \alpha_k \in \Bar{\alpha} \mid \alpha_k < \alpha_{v, th} \}| \geq p_v N,
\end{equation}
\begin{equation} \label{eq:cond2}
|\Bar{\alpha_e}| = |\{ \alpha_k \in \Bar{\alpha} \mid \alpha_k \geq \alpha_{e, th} \}| \geq p_e N
\end{equation}
where $\mid· \mid$ denotes the size of a subset, $\alpha_{v, th}$ and $\alpha_{e, th}$ are thresholds to ensure predominant Newtonian and predominant non-Newtonian regimes for some corresponding percentages, $p_v$ and $p_e$, of the total samples of the experiment, $N$. The conditions are nevertheless checked for a maximum number of iterations to ensure that, if the conditions are too difficult to match for a given parameters set, the simulator will not remain blocked in an endless loop. 

The input pressure signals are non-overlapping, randomly composed sequence of sine, ramp and step functions of different values generated according to a uniform distribution within the configured values. Each randomly generated pressure signal has a duration of $12.5s$ with a resolution of $0.05s$, or $250$ points at 20Hz. This pressure signal, which corresponds with $P_1$ in Figure \ref{fig:SysDiag}, remains constant while the dynamics of the rest of the system are calculated. Runge-Kutta \shortcite{rk1986} is used to integrate the state variable equation \eqref{eq:Vdot}. The used implementation assumes accuracy of order fourth to control the error, while steps are taken using the order five formulation. The generated dataset is composed of 5500 experiments of 12.5s (250 samples) each one. Figure \ref{fig:sim_ref} shows some of the main output signals from an example simulation.

\subsection{Artificial Neural Network}

\begin{figure}[t]
    \centering
    \includegraphics[width=0.45\textwidth]{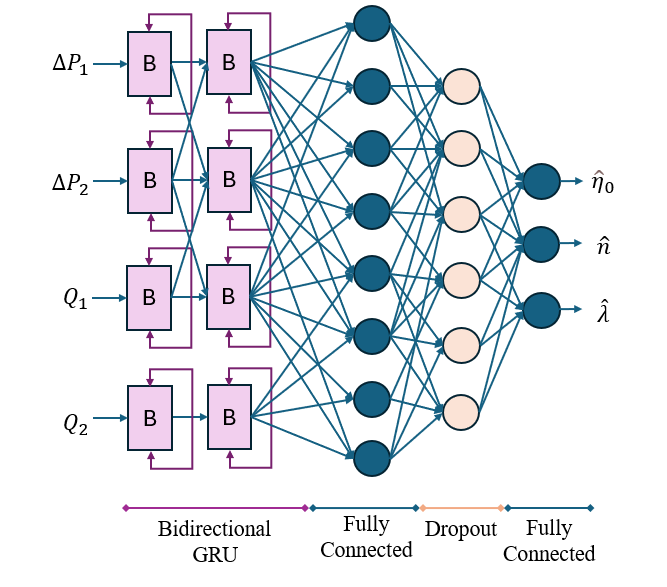}
    \caption{Designed artificial neural network. It receives four inputs signals: $\Delta P_1$, $\Delta P_2$, $Q_1$, $Q_2$, and outputs three viscosity parameters: $\eta_0$, $n$ and $\lambda$. The bidirectional GRU layers provides the model with memory capability to deal with temporal signals. The fully connected layers enhance its capacity to model complex functions. The dropout layer helps to regularize the model.}
    \label{fig:ann}
\end{figure}

The proposed architecture is composed of five layers, has four inputs (namely $\Delta P_1$, $\Delta P_2$, $Q_1$, $Q_2$) and three outputs (consisting of the estimated values of parameters $\eta_0$, $n$ and $\lambda$).

The two first layers of the model constitute the recurrent part, where different implementations were tested. Namely unidirectional Long Short-Term Memory (LSTM, \citeNP{lstm1997}), unidirectional Gated Recurrent Unit (GRU, \shortciteNP{gru2014}) and bidirectional GRU (BGRU, \shortciteNP{bgru2017}). In unidirectional RNN, information can only flow from past instants to the last instant, whereas in bidirectional RNN the signals are analyzed in both (time) directions \shortcite{brnn1997}. These BGRU layers were both implemented with twenty hidden units, Rectified Linear Units (RELU) as activation layers, leaky RELU for their recurrent activation, and a recurrent dropout of probability 10\%. 

After the two RNN layers, two fully connected (i.e. dense) layers are sequentially connected, separated by a dropout layer. The first input layer is implemented with twenty hidden units, followed by a leaky RELU activation layer. The dropout layer is set with probability 10\%. The output fully connected layer is composed by three units, one unit for each estimated parameter, with linear activation. The final model version has a total of 11.5k tunable parameters. Its architecture can be observed in Figure \ref{fig:ann}.

The inputs to the system are two signals of pressure drop and two signal of flow-rate. Each signal is individually normalized in range $(-1.0, 1.0)$, with a length of 250 samples per signal, 1000 samples in total. This configuration allows for its operation in real-time by implementing an input sliding window with the previous setup. The model produces only one estimate of each parameter, i.e., $\eta, n$, and $\lambda$, after inserting the 250 samples of each signal. The outputs are also in range $(-1.0, 1.0)$ and must be denormalized. The model was trained from zero. Adam optimizer is used with learning rate $2.5e-5$, no weight-decay, for 25 epochs, and Mean Squared Error (MSE) is used as loss function.

\subsection{Verification}

In order to verify the operation of the simulation, training, and estimation methodology, the following three steps procedure is performed: first, a K-fold cross validation method is applied to ensure the independence between the training and testing dataset. Thirty three percent of the generated dataset is used for the testing, and the remaining sixty seven percent is used for training. Thus, it is chosen $K=3$, performing a complete rotation over the dataset after three iterations of the method. Since the generation of the parameters is uniformly performed within the defined range and the dataset is shuffled before the execution of the K-fold cross validation, the performance is expected to be similar for the three iterations. To evaluate it, an accuracy normal distribution curve is fitted of each model. 

Afterwards, for each model, simulations are performed using the estimated parameters for each sample of the dataset. The Mean Square Error (MSE) between the curves simulated with the reference parameters and the curves simulated with the estimated parameters is calculated.

Finally, the Pearson correlation coefficients are calculated between the parameters estimation error and the simulated curves error. 
Correlation coefficients close to the unit imply a high correlation between these two error evaluation. This potentially indicates that the mapping between the parameters and the simulated curves is injective, and thus, the experiments are properly set for the training of the RNN.  Otherwise, a correlation coefficient for a particular estimated parameter closer to one half mean that the performed experiments might not be suitable for the RNN to discover the original values.

\section{EXPERIMENTS AND RESULTS} \label{s:ear}
In this section we explain how we have used the proposed methodology to explore the optimization of the simulation setup towards the inference of complex parameters, such as the rheological properties of a fluid, from pressure and flow-rate signals that do not have a straightforward linkage.

\begin{figure}[b]
    \centering
    \includegraphics[width=0.9\textwidth]{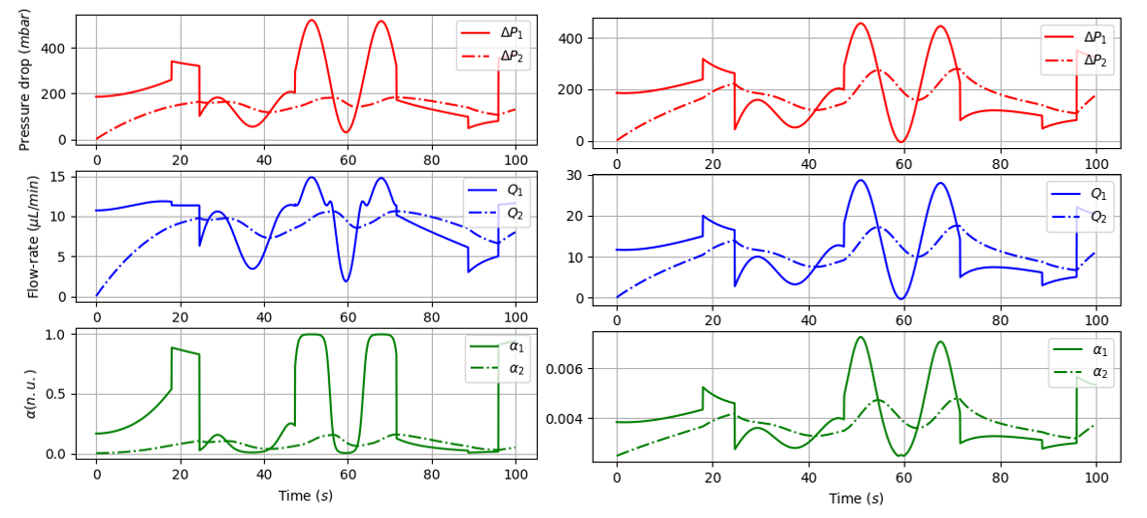}
    \caption{Two simulations performed over 100s with the same random pressure input signal. The parameters correspond to shear-thickening fluid ($n=1.1$) with, first, $\lambda = 1.0e-4$ (left), and second, $\lambda = 1.0e-5$ (right). In the rightmost simulation it can be observed that the fluid doesn't go into non-Newtonian regime ($\alpha \ll 1.0$), thus the viscosity remains low, producing higher-values of flow-rate. }
    \label{fig:sim_ref}
\end{figure}

In order to define the ranges over which the current setup could be useful, the proposed framework can be used to perform a heuristic analysis. For the given resistances, capacitance, and maximum input pressure (arbitrarily set to 600mbar or 60e3 Pa), the objective is to find out what are the ranges for the parameters of the viscosity model for which both Newtonian and non-Newtonian regimes can be observed. The most influential parameter in this regard is the relaxation time, $\lambda$, because it regulates the threshold flow-rate at which the fluid will behave in one or the other way. A higher zero-shear viscosity, $\eta_0$, will decrease the observed flow-rate, $Q_i$, for a given applied pressure, $P_1$, making it harder to reach the non-Newtonian regime. Figure \ref{fig:sim_ref} shows two example simulations with same random input signal. Non-Newtonian regime is only achieved in one of them. The resulting parameters' ranges from this analysis are found in Table \ref{tab:ranges}. 

\begin{figure}[t]
    \centering
    \includegraphics[width=1.0\textwidth]{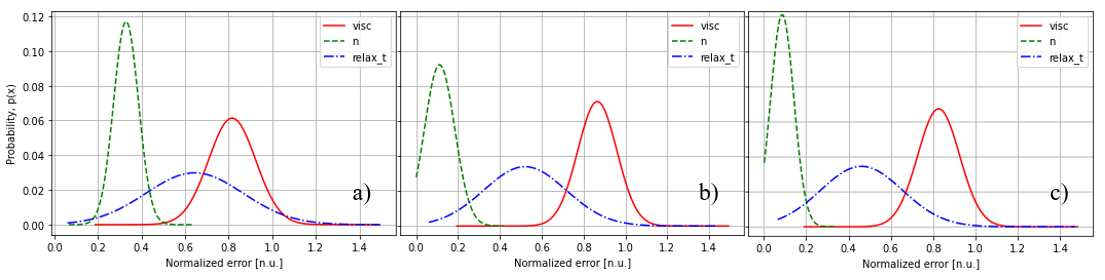}
    \caption{Error distributions for the three estimated parameters ($\eta_0$, $n$, and $\lambda$) training the deep learning model with three different random signal generation setups: a) only step sequences, b) steps and sinusoids sequences, and c) slower steps and sinusoids than in b). }
    \label{fig:error_dist}
\end{figure}

As a first step, an exploration of different RNN-based model designs was performed aiming to reduce the possibilities of having a performance bottleneck at the model architecture. Three RNN implementations were tested: unidirectional LSTM, unidirectional GRU and BGRU. While the benchmark was equivalent for all of them, the unidirectional LSTM often turned unstable, showing more sensitivity to the training hyper-parameters. Unidirectional GRU and BGRU implementations were both stable, being BGRU chosen since it was observed to be faster converging towards the same results for both same and different hyper-parameters.

After that, an exploration of the simulation setup is performed. Among the different parameters that can be configured are the properties of the randomized generation of pressure input signals: ranges for maximum and minimum values, ranges for the random frequency, the duration of each sequence inside the 10s experiment, etc. Figure \ref{fig:error_dist} shows the fitted error distributions for the estimation of each parameter of the viscosity model during the exploration process. 
Each distribution set of three distributions corresponds to a different deep learning model. The error distribution is normalized to ease their comparison. Each of these models has been trained with a dataset generated with a different configuration. In the first dataset, corresponding to Figure \ref{fig:error_dist}.a, the input pressure signal, $P_1$, is a sequence made out of 8-10 sinusoidal signals. In the second set, Figure \ref{fig:error_dist}.b, $P_1$ is a 8-10 segments sequence of steps, ramps, and sinusoidal signals. In the third one, Figure \ref{fig:error_dist}.c, 3-4 segments of steps, ramps and  signals are combined. The reduction of error can be observed, being the behavior exponent, $n$, the easiest parameter for the models to learn to estimate.

\begin{table}[b]
\centering
\caption{Pearson correlation coefficients between parameter estimation error vectors, ($E(\eta_0)$, $E(n)$, and $E(\lambda)$) and the simulation error vectors (E($\Delta P_1$), E($\Delta P_2$), E($Q_1$), and E($Q_2$)). The error vectors correspond to the model trained with input signals composed of 3-4 segments of steps, ramps, and signals. Associated error distributions are displayed at Figure \ref{fig:error_dist}.c. \label{tab:corr}}
\begin{tabular}{rllll}
\hline
 & E($\Delta P_1$) & E($\Delta P_2$) & E($Q_1$) & E($Q_2$) \\ \hline
$E(\eta_0)$ & $-0.183$ & $0.041$ & 0.101 & 0.117\\
$E(n)$ & 0.044 & \textbf{-0.533} & -0.066 & -0.164\\
$E(\lambda)$ & 0.107 & -0.054 & 0.076 & 0.095\\
\hline
\end{tabular}
\end{table}

The Pearson correlation coefficients are lastly calculated to be used as an indicator of sensitivity between the used input signals to the deep learning model and the parameters that it is asked to estimate. This is, if the chosen input signals to the model, $X$, are good candidates to estimate the parameters, $Y$, simulating the inaccurate values of the parameters, $\hat{Y}$, should produce wrong values of these input signals, $\hat{X}$. The error can be defined as $E(X)=\sqrt{X^2-\hat{X}^2}$. In Table \ref{tab:corr} the correlation coefficients are showed for the model whose results are displayed at Figure \ref{fig:error_dist}.c. The correlation shows that it is, indeed, the behavior index, $n$, the parameter with a higher correlation over the simulated signals, potentially making it the easiest parameter of viscosity model to be estimated.

\section{CONCLUSIONS AND FUTURE WORK} \label{s:cafw}

In this study, we developed a one-dimensional model for a hydraulic $RC$ circuit to simulate generalized Newtonian fluids. This model was crucial for generating synthetic data to train deep learning models. We used bidirectional GRUs in our methodology, which effectively predicted rheological parameters from the dynamic responses of microfluidic systems. Our findings show that the behavior index $n$ of the fluid correlates strongly with the output signals, making it the most reliably estimated parameter.

For future work, we plan to improve the simulation setup by automating the exploration process to optimize input signal configurations and model parameters systematically. We also aim to build a physical prototype to validate our simulations and refine model assumptions. This step is essential for practical applications in industrial and medical diagnostics. Additionally, we will explore incorporating more complex fluid dynamic models and advanced neural network architectures to handle a wider range of fluids and flow conditions, expanding the applicability of our methodology to various industrial uses.

\vspace{-0.2cm}
\section*{ACKNOWLEDGMENTS}
This work has been supported by the Research Projects IA-GESBLOOM-CM (Y2020/TCS-6420) funded by the Synergic program of the Comunidad Autónoma de Madrid (CAM), SMART-BLOOMS (TED2021-130123B-I00) by MCIN/AEI/10.13039/501100011033 and the European Union by NextGenerationEU/PRTR, and the MSCA-ITN-H2020 Project LasIonDef (GA n.956387).

\vspace{-0.3cm}
% Reducing font size (to 9pt) for References & Author Biagraphies
\footnotesize

% Please don't exchange the bibliographystyle style
\bibliographystyle{wsc}

% AUTHOR: Include your bib file here
\bibliography{references}

\section*{AUTHOR BIOGRAPHIES}

\noindent {\bf \MakeUppercase{Juan Sandubete-López}} is an industrial PhD candidate by Universidad Complutense de Madrid, and at Microfluidic Innovation Center, Paris. He holds a masters degree in Systems and Control Engineering from the same university. His research interests include autonomous systems, instrumentation, deep learning, and simulation. His email address is \email{jsandu01@ucm.es}.\\

\noindent{\bf\uppercase{José L. Risco-Martín}} received his Ph.D. from UCM, where he currently is Full Professor in the Department of Computer Architecture and Automation. His research interests include systems modeling, simulation, and optimization. His email address is \email{jlrisco@ucm.es}.\\

\noindent {\bf \MakeUppercase{Alexander McMillan}} is an R\&D project leader at the Microfluidics Innovation Center, where he oversees a team that develops new microfluidic tools through academia-industry collaborations. He earned his PhD from KU Leuven School of Bioscience Engineering, focusing on new polymers for microfluidic devices and their applications in flow chemistry and cell culture. His email address is \email{alex.mcmillan@microfluidic.fr}.\\

\noindent{\bf\uppercase{Eva Besada-Portas}} is an Associate Professor of Systems Engineering and Automation at UCM. She also holds a PhD in Computer Systems from UCM. Her research interests include uncertainty modeling and simulation, optimal control and planning of unmanned vehicles. Her email address is \email{ebesada@ucm.es}.\\

\end{document}

%% file: wscbib.tex
%%%%%%%%%%%%%%%%%%%%%%%%%%%%%%%%%%%%%%%%%%%%%%%%%%%%%%%%%%%%%%%%%%%%%%%%%%%%%%
%                                                                            %
%     THESE COMMANDS ARE REQUIRED TO WORK WITH WSC.BST TO MAKE BIBLIO     %
%                                                                            %
%%%%%%%%%%%%%%%%%%%%%%%%%%%%%%%%%%%%%%%%%%%%%%%%%%%%%%%%%%%%%%%%%%%%%%%%%%%%%%
\makeatletter
\let\@internalcite\cite
\def\cite{\def\@citeseppen{-1000}%
    \def\@cite##1##2{(##1\if@tempswa , ##2\fi)}%
    \def\citeauthoryear##1##2##3{##1 ##3}\@internalcite}
\def\citeNP{\def\@citeseppen{-1000}%
    \def\@cite##1##2{##1\if@tempswa , ##2\fi}%
    \def\citeauthoryear##1##2##3{##1 ##3}\@internalcite}
\def\citeN{\def\@citeseppen{-1000}%
%  Pierre L'Ecuyer's fix for multiple cite bug
%  Added by Paul J Sanchez on 4 October 2001
%   \def\@cite##1##2{##1\if@tempswa , ##2)\else{)}\fi}%
%   \def\citeauthoryear##1##2##3{##1 (##3}\@citedata}
    \def\@cite##1##2{##1\if@tempswa, ##2)\else{}\fi}%
    \def\citeauthoryear##1##2##3{##1 (##3)}\@citedata}
\def\citeA{\def\@citeseppen{-1000}%
    \def\@cite##1##2{(##1\if@tempswa , ##2\fi)}%
    \def\citeauthoryear##1##2##3{##1}\@internalcite}
\def\citeANP{\def\@citeseppen{-1000}%
    \def\@cite##1##2{##1\if@tempswa , ##2\fi}%
    \def\citeauthoryear##1##2##3{##1}\@internalcite}
\def\shortcite{\def\@citeseppen{-1000}%
    \def\@cite##1##2{(##1\if@tempswa , ##2\fi)}%
    \def\citeauthoryear##1##2##3{##2 ##3}\@internalcite}
\def\shortciteNP{\def\@citeseppen{-1000}%
    \def\@cite##1##2{##1\if@tempswa , ##2\fi}%
    \def\citeauthoryear##1##2##3{##2 ##3}\@internalcite}
\def\shortciteN{\def\@citeseppen{-1000}%
%  Pierre L'Ecuyer's fix for multiple cite bug
%  Added by Paul J Sanchez on 2 September 2002
%  should have caught this last year...
%   \def\@cite##1##2{##1\if@tempswa , ##2)\else{)}\fi}%
%   \def\citeauthoryear##1##2##3{##2 (##3}\@citedata}
% Shane G. Henderson fix for extra right bracket at end of optional material June 8, 2005
%    \def\@cite##1##2{##1\if@tempswa, ##2)\else{}\fi}%
    \def\@cite##1##2{##1\if@tempswa, ##2\else{}\fi}%
    \def\citeauthoryear##1##2##3{##2 (##3)}\@citedata}
\def\shortciteA{\def\@citeseppen{-1000}%
    \def\@cite##1##2{(##1\if@tempswa , ##2\fi)}%
    \def\citeauthoryear##1##2##3{##2}\@internalcite}
\def\shortciteANP{\def\@citeseppen{-1000}%
    \def\@cite##1##2{##1\if@tempswa , ##2\fi}%
    \def\citeauthoryear##1##2##3{##2}\@internalcite}
\def\citeyear{\def\@citeseppen{-1000}%
    \def\@cite##1##2{(##1\if@tempswa , ##2\fi)}%
    \def\citeauthoryear##1##2##3{##3}\@citedata}
\def\citeyearNP{\def\@citeseppen{-1000}%
    \def\@cite##1##2{##1\if@tempswa , ##2\fi}%
    \def\citeauthoryear##1##2##3{##3}\@citedata}
%
% \@citedata and \@citedatax:
%
% Place commas in-between citations in the same \citeyear, \citeyearNP,
% \citeN, or \shortciteN command.
% Use something like \citeN{ref1,ref2,ref3} and \citeN{ref4} for a list.
%
\def\@citedata{%
    \@ifnextchar [{\@tempswatrue\@citedatax}%
                  {\@tempswafalse\@citedatax[]}%
}

\def\@citedatax[#1]#2{%
\if@filesw\immediate\write\@auxout{\string\citation{#2}}\fi%
  \def\@citea{}\@cite{\@for\@citeb:=#2\do%
    {\@citea\def\@citea{, }\@ifundefined% by Young
       {b@\@citeb}{{\bf ?}%
       \@warning{Citation `\@citeb' on page \thepage \space undefined}}%
{\csname b@\@citeb\endcsname}}}{#1}}%

% don't box citations, separate with ; and a space
% also, make the penalty between citations negative: a good place to break.
%
\def\@citex[#1]#2{%
\if@filesw\immediate\write\@auxout{\string\citation{#2}}\fi%
  \def\@citea{}\@cite{\@for\@citeb:=#2\do%
    {\@citea\def\@citea{; }\@ifundefined% by Young
       {b@\@citeb}{{\bf ?}%
       \@warning{Citation `\@citeb' on page \thepage \space undefined}}%
{\csname b@\@citeb\endcsname}}}{#1}}%

% (from apalike.sty)
% No labels in the bibliography.
%
\def\@biblabel#1{}
\makeatother

%\newlength{\bibhang}
%\setlength{\bibhang}{2em}

% Indent second and subsequent lines of bibliographic entries. Taken
% from openbib.sty: \newblock is set to {}.
% \renewcommand{\refname}{REFERENCES}

\newdimen\bibindent
\bibindent=0.0em
% SEC: was \def\thebibliography#1{\section*{\refname\@mkboth
% SEC: was   {\uppercase{\refname}}{\uppercase{\refname}}}\list
\def\thebibliography#1{\section*{\refname}\list
   {}{\settowidth\labelwidth{[#1]}
   \leftmargin\parindent
   \itemindent -\parindent
   \listparindent \itemindent
   \itemsep 0pt
   \parsep 0pt}
   \def\newblock{}
   \sloppy
   \sfcode`\.=1000\relax}